\documentclass[10pt]{article}
\usepackage[OE]{express}
\usepackage{cite}
\begin{document}

\title{Engineering the axial intensity of Bessel beams}

\author{Raghu Dharmavarapu,\authormark{1,2,*} Saulius Juodkazis,\authormark{2,3} and \\\protect Shanti Bhattacharya\authormark{1}}

\address{\authormark{1}Centre for NEMS and Nanophotonics (CNNP), Department of Electrical Engineering, Indian Institute of Technology Madras, Chennai 600036, India\\
\authormark{2}Centre for Micro-Photonics, Faculty of Science, Engineering and Technology, Swinburne University of Technology, Hawthorn VIC 3122, Australia\\
\authormark{3}{Melbourne Centre for Nanofabrication, the Victorian Node of the Australian National Fabrication Facility, 151 Wellington Rd., Clayton 3168 VIC, Australia}}
\email{\authormark{*}raghu.d@ee.iitm.ac.in} 


\begin{abstract}
Bessel beams (BBs) appear immune to diffraction over finite propagation distances due to the conical nature of light propagation along the optical axis. This offers promising advantages in laser fabrication. However, BBs exhibit a significant intensity variation along the direction of propagation. We present a simple technique to engineer the axial intensity of the BB over centimeters-long propagation distances without expansion of the incoming laser beam. This method uses two diffractive optical elements (DOEs), one converts the input Gaussian intensity profile to an intermediate intensity distribution, which illuminates the second DOE,  a binary axicon. BBs of desired axial intensity distribution over few centimeters length can be fabricated. 
\end{abstract}


\section{Introduction}

Bessel beams (BB) have an elongated axial light intensity distribution that is immune to diffraction over finite propagation distances. The analytical expression of the BB intensity~\cite{sheppard1978gaussian} was refined in ref.~\cite{durnin1987diffraction}, where it was demonstrated that the zeroth order BB is a member of a special class of solutions to the Helmholtz equation that are propagation invariant. 

The transverse intensity profile of the BB follows a zeroth order Bessel function, which has a high intensity central peak surrounded by number of concentric rings. Unlike a traditional Gaussian beam, whose beam waist diverges, the transverse intensity profile of the BB remains unchanged as it propagates. The ideal BB with rings extending infinitely in the radial direction maintains constant axial intensity (an infinite amount of energy). Therefore, only radially truncated approximations of BBs are realized. Bessel beams also exhibit another interesting property known as self-healing, i.e., the ability of the beam to reconstruct after encountering an obstacle. Owing to these special properties, these beams are being used as an alternative to Gaussian beams in many applications such as optical manipulation~\cite{arlt2001optical}, high precision hole drilling~\cite{matsuoka2006characteristics} and light sheet microscopy~\cite{planchon2011rapid}. Several techniques have been proposed to generate these beams: refractive axicons~\cite{mcleod1954axicon}, diffractive axicons~\cite{dyson1958circular} and annular apertures~\cite{airy1841diffraction}.

Even though the above mentioned techniques can be used to generate Bessel-like beams, they all suffer from non-uniformity of intensity along the axis. The axial intensity variations along the  axis limits applicability of these beams. Several techniques have already been reported to create BBs with specific axial intensity profiles~\cite{Cizmar2009,ouadghiri2016arbitrary}. An axicon with an annular aperture and a logarithmic phase profile has been proposed to flatten~\cite{jaroszewicz1993apodized} the axial intensity of the Bessel-like beams. It was shown that by superposing a finite number of BBs with different longitudinal wave numbers, one can produce arbitrary desired axial intensities known as ``frozen waves''~\cite{zamboni2004stationary}. For emerging industrial applications of laser dicing, drilling, and inscription of optical elements, an axial control of beam intensity and axially moving focus actuated by a transverse acoustical wave in the lens~\cite{shih2012two} are among actively researched applications. Using BBs for material modification~\cite{marcinkevivcius2001application,mikutis2013high}, it is preferable to have simple optical elements, which can be directly inserted into the laser's Gaussian-like beam to generate BBs.

Here, we demonstrate a technique that uses two diffractive optical elements (DOEs), where the first DOE converts the incident Gaussian beam into an intermediate intensity distribution which illuminates the second DOE, a diffractive axicon, that produces the desired on-axis intensity. In this manner, the final axial intensity can be engineered to have any desired variation by suitably adjusting the intensity output of the first DOE. This is demonstrated with examples of: (i) a linearly increasing, (ii) an uniform, and (iii) exponentially increasing axial intensity distributions.  The DOEs were designed and the axial intensity outputs were simulated. Simulation results were compared with experimental results for the case (i). The DOEs were fabricated using electron beam lithography (EBL) and desired on-axis intensity profiles were achieved over long propagation distances of few centimeters.

\section{Theory: Bessel  beam}

\begin{figure}[tb]
\centering
\includegraphics[width=11.5cm]{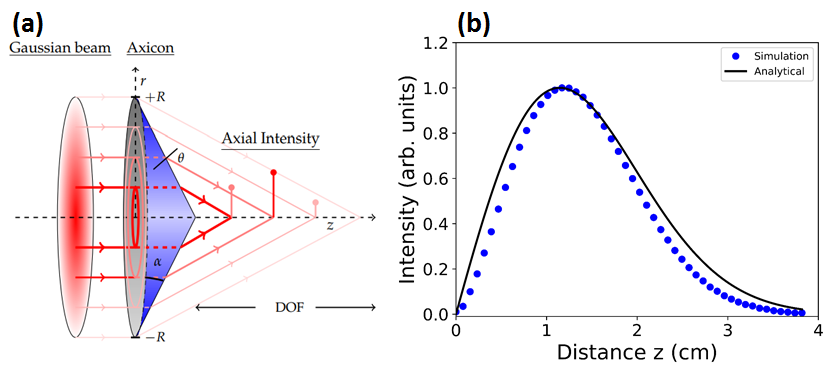}
\caption{(a) Refractive axicon ray tracing for the Gaussian input; DOF is the depth of focus. (b) Analytical (line) and Fresnel integral simulated (dots) axial intensity of the Bessel beam generated by an axicon with radius $R = 2$~mm and $\alpha = 1.6^\circ$.}
\label{fig:1}
\end{figure}

Axicons - conical lenses - are optical elements that have rotational symmetry about the $z$-axis (Fig.~\ref{fig:1}). They generate a quasi-Bessel beam throughout their depth of focus (DOF) region. Beyond the DOF, the beam gradually transforms into a ring of constant width and increasing the radius as it propagates. The important parameters that characterize an axicon are its front face radius, $R$, the cone angle, $\alpha$, and the refractive index, $n$ (Fig.~\ref{fig:1}). These parameters together determine the length of the DOF of the axicon. 

Consider the input light beam to be a collection of rays traveling parallel to the $z$-axis. All these rays refract at the conical surface of the axicon towards the axis with the same angle $\theta$. All the rays at one radial distance, come to focus at one point on the axis. The rays incident at the extreme of the axicon (i.e., the furtherest radial distance) determine the DOF of the axicon, as shown in Fig.~\ref{fig:1}. Using Snell's law, $\sin\theta = n\sin\alpha$ and applying the small angle approximation for $\alpha$, the DOF is found to be:
\begin{equation}
DOF = \frac{R}{(n-1)\alpha}.
\label{eq:1}
\end{equation}

Even though the axicon generates a BB, the intensity along the axis is not uniform within the DOF. To demonstrate this, we consider a standard Gaussian intensity profile for the incident beam, i.e.,
\begin{equation}
I_{in}(r) = I_0 \exp{\left(\frac{-2r^2}{w_0^2}\right)},
\end{equation}
where $w_0$ is the beam waist (radius), $I_0$ is the peak intensity of the Gaussian beam and $r=\sqrt{x^2+y^2}$ is the radial distance from the $z$-axis. To derive an analytical expression for the on-axis intensity, we consider a thin annular ring of width $dr$, inner radius $r$ and outer radius $r+dr$ on the front surface of the axicon. The amount of power passing through this annular ring is given by:
\begin{align}\label{eq:4}
P_{ring}(r) &= I_0\exp{\left(\frac{-2r^2}{w_0^2}\right)}[\pi(r+dr)^2)-\pi r^2]
     \approx I_0\exp{\left(\frac{-2r^2}{w_0^2}\right)}2\pi r dr.
\end{align}
This power gets spread out along a length $dz$ on the $z$-axis. Therefore, the axial intensity is given by $AI(z) = P_{ring}(r)/dz$. Dividing both sides of Eq.~\ref{eq:4} with $dz$ and substituting $r = z(n-1)\alpha$ from Eq.~\ref{eq:1} gives:
\begin{align}
AI(z) = I_0\exp{\left(\frac{-z^2(n-1)^2\alpha^2}{w_0^2}\right)}2\pi z(n-1)^2\alpha^2
\label{eq:5}
\end{align}
\noindent From Eq.~\ref{eq:5}, it can be seen that the on-axis intensity of a BB generated using an axicon is directly related to the input intensity profile $I_{in}(r)$. Equation~\ref{eq:5} can be generalized for any arbitrary input, $I_{in}$, as: 
\begin{align}
AI(z) = MI_{in}(z(n-1)\alpha)z,
\label{eq:6}
\end{align}
where $M = 2\pi (n-1)^2\alpha^2$ is a constant for the designed axicon. It should be noted that Eq.~\ref{eq:6} should be used to compute the on-axis intensity only for $0<z<$DOF. 

To confirm the validity of this equation, we have compared the axial intensity derived from Eq.~\ref{eq:6} with the Fresnel simulations of a diffractive axicon for a Gaussian input shown in Fig.~\ref{fig:1}(b). The axicon parameters used were: $R = 1$~mm and $\alpha = 1.6^\circ$, which results in a DOF of 3.5~cm useful for practical applications. Simulated results were in good agreement with analytical calculations.

\section{Samples and fabrication}

The DOEs were designed with a diameter of 2~mm operating at the wavelength of 633~nm. Polymethylmethacrylate (PMMA) resist 950~K A8 (MicroChem GmbH) was used as the electron beam lithography (EBL) resist. Indium tin oxide (ITO) coated glass plate was used as a substrate. The ITO layer prevents charging during EBL writing. Presence of the ITO layer decreased the transmission to $85\%$ for 633~nm. The EBL parameters were as follows: acceleration voltage 10~kV, aperture $30~\mu$m and dose $70~\mu$C/cm$^2$. The DOEs were developed in a mixture of methyl isobutyl ketone (MIBK) and isopropyl alcohol (IPA) at ratio 1:3 for 50~s followed by cleaning in IPA for 30~S.

\begin{figure}[tb]
\centering
\includegraphics[width=12cm]{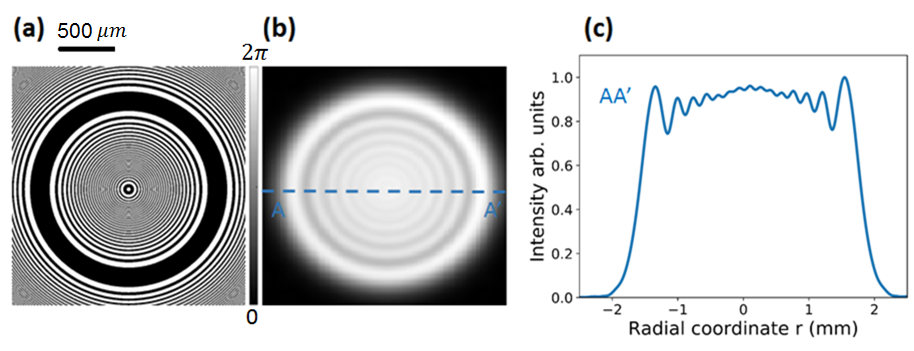}
\caption{Simulation. The phase of DOE1 (a) and  intensity (b) profiles created at the front surface of axicon $10~cm$ behind the DOE1. The phase span in (a) is $0-2\pi$. (c) Intensity profile along the central line \emph{AA'} shown in (b).}
\label{fig:Exp1}
\end{figure}

\begin{figure}[tb]
\centering
\includegraphics[width=6.5cm]{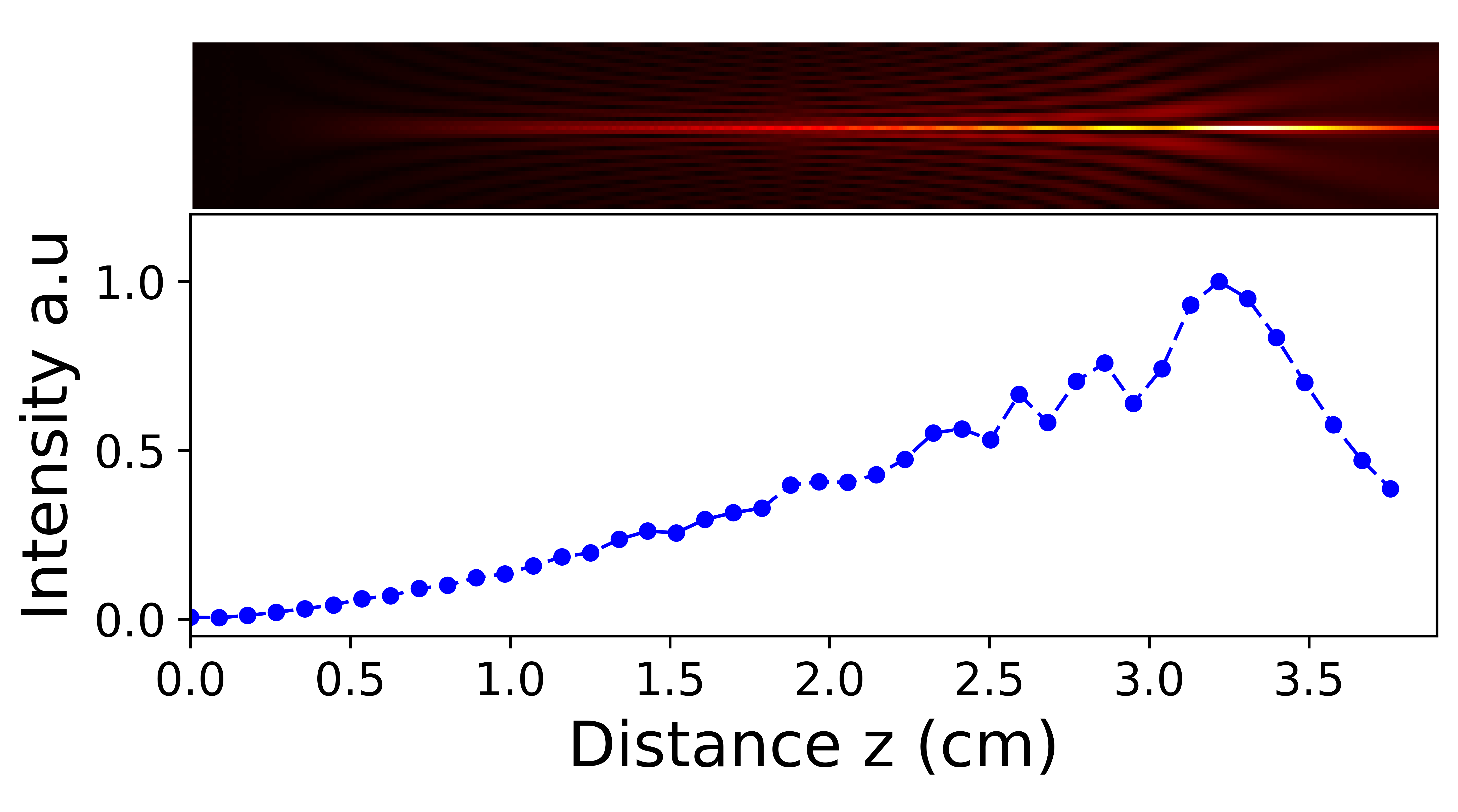}
\caption{Fresnel simulation (same method as in Fig.~\ref{fig:1}) of linearly increasing axial intensity and beam cross section in $yz$-plane.}
\label{fig:LIAISim}
\end{figure}

\section{Results and Discussion}

The relationship between the input (lateral) intensity to the diffractive axicon (DOE2) and the axial intensity was derived by using the geometrical law of energy conservation. DOE1 transforms a Gaussian beam into the desired intermediate intensity profile at the plane of the diffractive axicon. Its phase profile is calculated using the simplified mesh technique~\cite{bhattacharya2008simplified} or the Gechberg-Saxton (G-S) algorithm~\cite{gerchberg1972practical}. The output is simulated by calculating the intensities at different planes along the beam propagation direction using the Fresnel diffraction integrals~\cite{goodman2005introduction}. The DOEs were fabricated using EBL. With this method, desired on-axis intensity profiles were achieved over long propagation distances of few centimeters.

\begin{figure}[tb]
\centering
6\includegraphics[width=12cm]{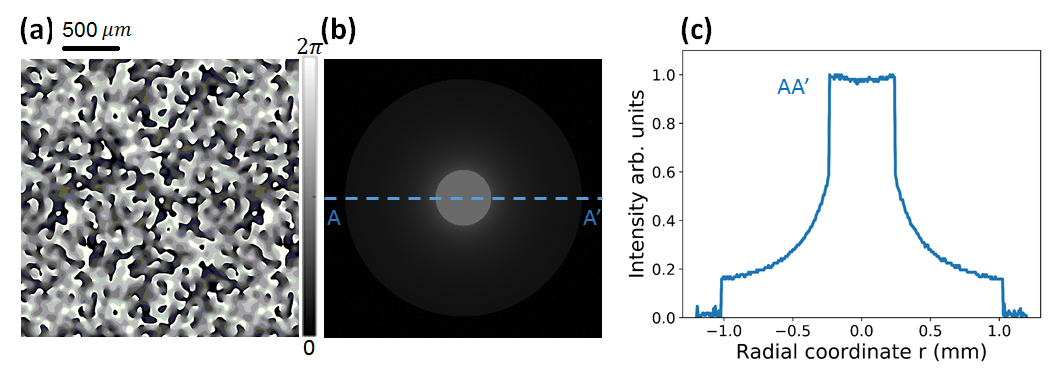}
\caption{Simulation results. (a) Phase profile of DOE1 with random phase, which is a standard feature of a pure phase distribution generated by the G-S algorithm. This pattern is more complex to fabricate and it is more susceptible to scattering.  (b) Intensity profile created at the front surface of axicon, 10~cm behind DOE1. (c) Intensity profile along the central line \emph{AA'} cross section shown in (b).} 
\label{fig:Exp2}
\end{figure}	

\subsection{Engineering the axial intensity}

For an axicon, the on-axis intensity rises to a peak and gradually declines towards the end of the DOF (Fig.~\ref{fig:1}(a)),  when the incident beam is Gaussian-like. Therefore, one can only utilize these beams over a very limited axial region, where the intensity might be considered as uniform.  Alternately, the beams could be used in applications where uniformity is not important. As it was shown, the the on-axis intensity is directly linked to the incident intensity profile through Eq.~\ref{eq:6}. Therefore, this relationship can be used to back calculate the input intensity $I_{in}(r)$ that gives the desired $AI(z)$. As most laser sources will have a Gaussian intensity profile, we need to design an optical element DOE1 that coverts the Gaussian to the desired $I_{in}(r)$.

\begin{figure}[tb]
\centering
\includegraphics[width=14cm]{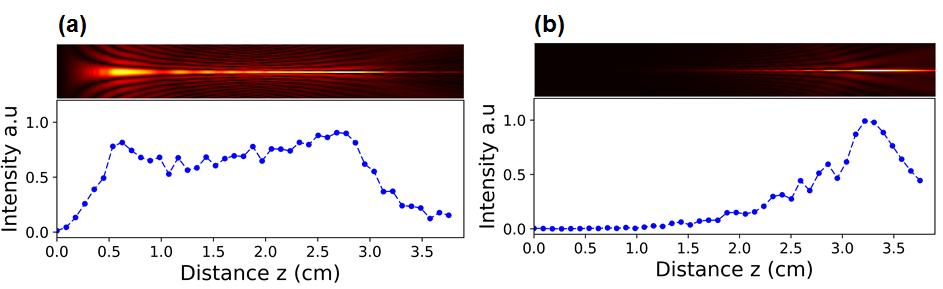}
\caption{Fresnel simulation of the Bessel beam. (a) Uniform axial intensity, (b) parabolically increasing axial intensity with beam cross section in $yz$-plane shown above.}
\label{fig:UAISim}
\end{figure}

A technique to engineer the axial intensity using two DOEs is proposed next. The first DOE transforms the Gaussian beam into an intermediate intensity $I_{in}(r)$ at the plane of the second DOE, which is a diffractive axicon. We have used simplified mesh technique\cite{bhattacharya2008simplified} to compute the phase profile ($\phi(x,y)$) of DOE1. 

\subsubsection{Simulation: a linearly increasing axial intensity}

For a linearly increasing axial intensity, the axial intensity is:
\begin{align}
AI(z) = az, 
\end{align}
where $a$ is a positive constant. We get the desired input intensity for the diffractive axicon, by substituting this in Eq.~\ref{eq:6}:
\begin{align}
I_{in}(r) = \frac{a}{M},
\end{align}
i.e., a flat-top beam. Since the input and the desired output intensities are known, they can be used to calculate the phase distribution that would convert one to the other. While the G-S algorithm is perfect for such problems, it generates a random phase variation, which leads to scattering in practical systems. In order to avoid this, the calculated phase distribution should be continuous.  

In the simplified mesh technique, the incident and output beams are each divided into a mesh consisting of  zones of equal power. Eikonal equations~\cite{born1999principles} are used to connect the input to the output zones. The phase distribution $\phi(x,y)$, required to produce the desired output is obtained by solving these equations~\cite{bhattacharya2008simplified,hermerschmidt1998design}.

\begin{figure}[tb]
\centering
\includegraphics[width=8.5cm]{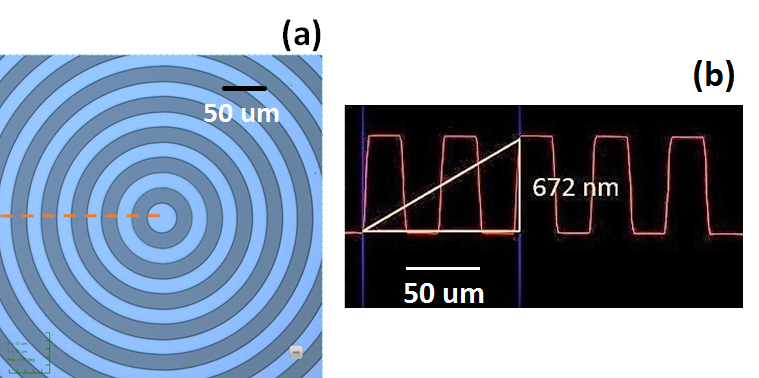}
\caption{(a) Microscope images of fabricated diffractive axicon designed for $R = 1$~mm and $DOF = 3.5$~cm. (b) Depth profile measured with confocal microscope.}
\label{fig:fab}
\end{figure}

Since the energy in each zone in the input plane is directed to a similarly located zone in the output plane and as the equations are solved simultaneously, the phase obtained is continuous. It should be pointed out that this technique is useful in cases, where the meshes are easy to construct. 
 
The input Gaussian beam waist was chosen as 1.1~mm to match the laser beam  used in the experiment. DOE1 creates a circular flat-top beam.  Simulated $I_{in}(r)$ at the input of DOE2 and the on-axis intensity are shown in Figs.~\ref{fig:Exp1} and \ref{fig:LIAISim}, respectively. 

\begin{figure}[tb]
\centering
\includegraphics[width=9.7cm]{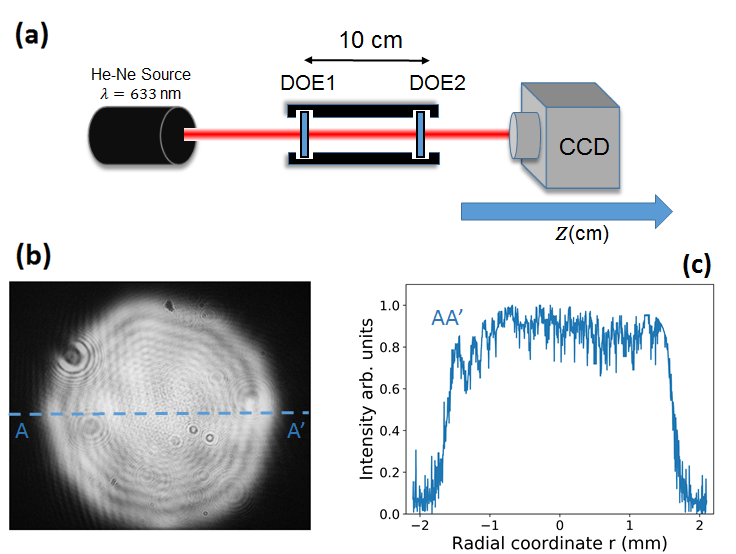}
\caption{(a) Experimental setup. (b) Flat top intensity profile. (c) Intensity plotted along the central line of (b).}
\label{fig:Flat}
\end{figure}

\subsubsection{Simulation: an uniform axial intensity}

For many practical applications the desired is the uniform axial intensity is generated using:
\begin{align}
\begin{split}
AI(z) &= az \quad \mathrm{ for }\quad 0<z<d_1;\\
&=a \quad \mathrm{ for }\quad d_1<z<DOF;
\end{split}
\end{align}
where $a$ is a positive constant. The on-axis intensity is constant between the axial points $d_1$ and the depth of focus $DOF$. The desired $I_{in}(r)$ to create this axial intensity is:
\begin{align}
 I_{in}(r) = \frac{(n-1)\alpha}{M} \frac{1}{r},
\end{align}
which is a hyperbolic intensity profile. In this case, the phase profile $\phi(x,y)$ that transforms the input Gaussian beam into into the desired intensity profile was computed using the G-S algorithm. This is because the mesh of the output beam was more complicated for this intensity distribution. The corresponding Fresnel simulations are shown in Fig.~\ref{fig:Exp2}. Simulated on-axis intensity is presented in Fig.~\ref{fig:UAISim}(a). Such intensity distribution when the high intensity is not located at the very tip of axicon ($z = 0$) but is transfered to the axial position $z = d_1$ is very useful for high-power laser cutting and welding applications due to reduction of optical dama.   

\subsubsection{Simulation: a parabolic axial intensity}

\begin{figure}[tb]
\centering
\includegraphics[width=7.50cm]{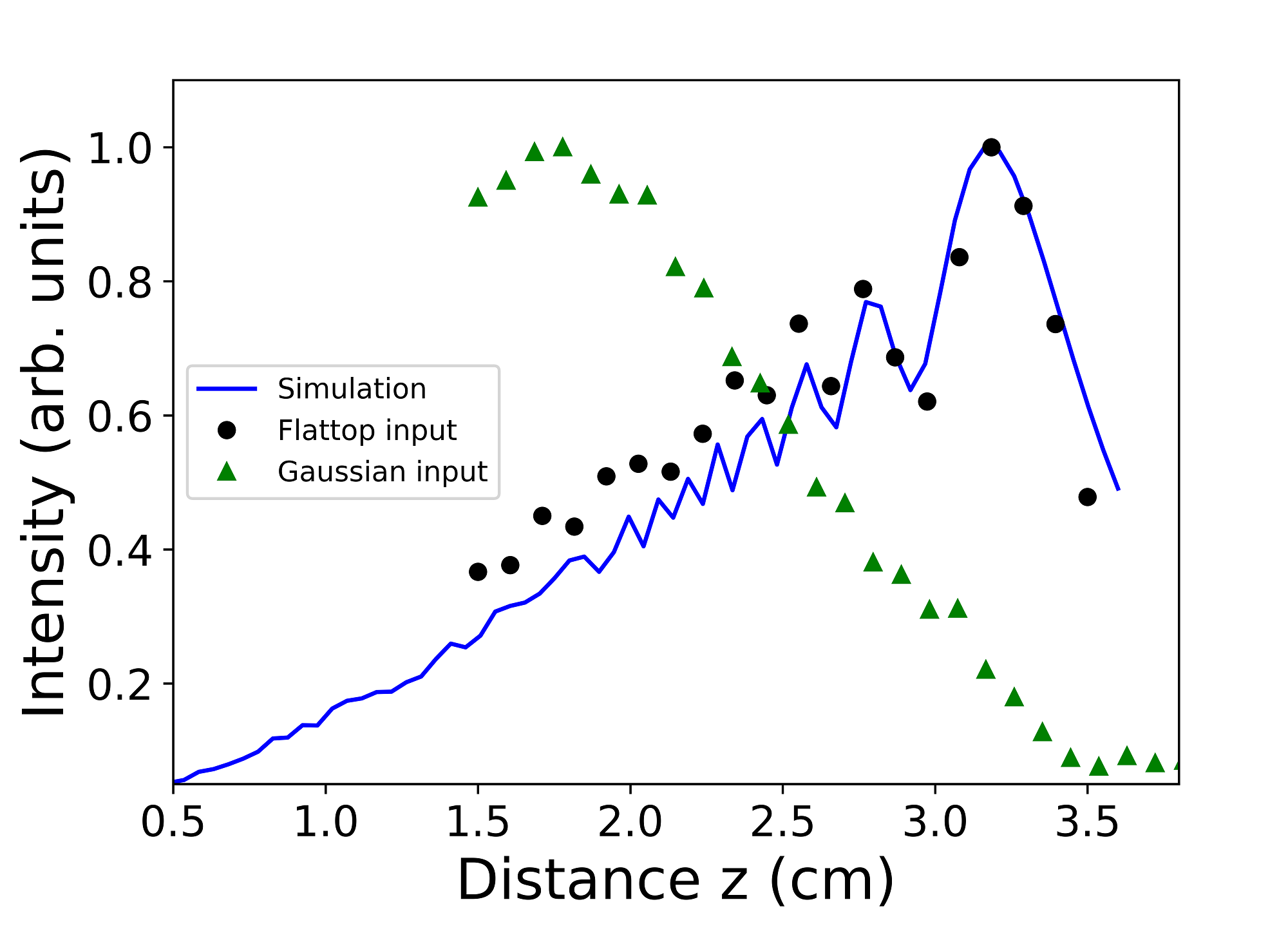}
\caption{On-axis intensity variation after DOE2. The solid line shows the simulated and dotted circles show the experimentally measured axial intensity when the flat-top beam was incident on DOE2. The dotted triangles curve shows the experimental results for an incident Gaussian beam.}
\label{fig:Exp}
\end{figure}

Finally, a parabolic axial intensity profile can be realized:
\begin{align}
AI(z) = az^2,
\end{align}
where $a$ is a positive constant. The desired $I_{in}(r)$ to get this axial intensity is,
\begin{align}
I_{in}(r) = \frac{az}{M},
\end{align}
which is a radially increasing intensity distribution. The simulated axial intensity is shown in Fig.~\ref{fig:UAISim}(b).

\subsection{Fabricated Bessel beam generator}

The experimental setup used for characterisation of the fabricated optical elements is shown in Fig.~\ref{fig:Flat}(a). A collimated beam from a He-Ne source with $1/e^2$ diameter of 1.1~mm was passed through the DOE1. This DOE1 was designed to create the desired $I_{in}(r)$  at the distance of 10~cm behind the element, at the location of DOE2 (such alignment of elements is suitable for practical application in laser structuring). Care has been taken to align the center of both DOE1 and DOE2. The on-axis intensity was measured by moving a CCD camera along the axis using a translation stage.
 
DOE1 and DOE2 were fabricated using a Raith 150-Two EBL system. The confocal microscope images of the diffractive axicon and its profile are shown in Fig.~\ref{fig:fab}.

The experimental results of the Gaussian to flat-top beam conversion with DOE1 are shown in Fig.~\ref{fig:Flat}(b). Measured values of the on-axis intensity  for the Gaussian and flat-top incidence are summarised in Fig.~\ref{fig:Exp}. The axial intensity within the first 1.5~cm could not be measured as the CCD sensor housing had a depth of $\approx 1.5$~cm within the C-mount.


We used two binary DOEs written on ITO coated borosilicate glass substrate in tandem to achieve desired axial intensity profile. The theoretical maximum transmittance of each of these DOEs is 81\% at the used wavelength of 632~nm which results in an over all efficiency of 65\%. Further reduction in efficiency due to diffraction losses is a drawback of this technique as the DOEs are binary in nature. This situation can be improved by fabricating multilevel 3D DOEs using advanced fabrication techniques such as a grayscale EBL, scanning tip method for 3D structuring of resist, e.g., using NanoFrazor, or polymerization via direct laser writing~\cite{polim}. Flat meta-optical elements~\cite{khorasaninejad2017metalenses} can also be used to fabricate these DOEs, which offer high resolution spatially varying phase that can minimize diffraction losses. Higher efficiencies could be achieved by depositing an anti-reflection coating on the DOEs. Since the purpose of DOE1 is only to generate an intermediate $I_{in}(r)$ for a predetermined axial-intensity and does not essential for generation of the BB, the DOF ranging from few millimeters-to-centimeters can be tuned just by changing the DOE2. 

Typically it is difficult to manufacture refractive axicons with cone angles less than $\sim 1^\circ$ which puts a limitation on the length of few millimeters of DOF. With proposed technique longer DOFs up to centimeters are possible as the cone angle is related to the number of rings of the diffractive axicon, which can be easily fabricated with available micro/nano-lithography tools. As this technique does not use any spatial light modulators for beam shaping, the two compact DOEs can be directly incorporated easily into any practical laser fabrication or imaging application.

\section{Conclusions}

A simple method to engineer the axial intensity of the BB, using two simple DOEs, over longer lengths 3.5~cm is demonstrated. To the best of our knowledge this is the first time, engineering of the beam intensity has been done over such large distances. This method is particularly advantageous over methods that use spatial light modulators, as those devices are expensive and difficult to include in industrial applications. We have designed DOEs to generate BBs with a linearly increasing axial intensity and with the uniform axial intensity. Experimental results show a good agreement with simulations. 

\section*{Funding}

Partial support via the Australian Research Council DP170100131 Discovery project.

\section*{Acknowledgments}
The authors thank Center for NEMS and Nanophotonics (CNNP) for the use of their fabrication facilities.
\bibliography{bib}
\end{document}